\documentclass{llncs}

\usepackage{amssymb}
\usepackage{amsmath}
\usepackage{amsfonts}
\usepackage{amsxtra}
\usepackage{amscd}
\usepackage{epsfig}
\usepackage{enumerate}
\usepackage{float}
\usepackage{xy}

\xyoption{all}

\newcommand{\Gal}{\mbox{$\mathrm{Gal}$}}

\newcommand{\GL}{\mbox{$\mathrm{GL}$}}

\newcommand{\newfontobj}[2]{
  \newcommand{#1}[1]{
    \expandafter\def\csname##1\endcsname{{#2 ##1}}}}

\newfontobj{\class}{\rm}
\newfontobj{\lang}{\bf}
\newfontobj{\oper}{\rm}

\class{PSPACE}		
\class{AM}		
\class{MA}
\class{NP}
\class{UP}
\class{P}
\class{RP}
\class{BPP}
\class{DTIME}
\class{ZPP}
\class{EXPSPACE}
\class{coNP}
\class{coRP}
\class{coAM}
\class{PH}
\class{co}
\class{BP}	

\lang{HN} 		
\lang{SAT}
\lang{PIM}
\lang{CSG}
\lang{ESC}
\lang{IP}
\lang{PRIMES}

\oper{Pr}    
\oper{E}     
\oper{Ord}   
\oper{li}

\floatstyle{ruled}
\newfloat{Algorithm}{thp}{loa}[section]

\usepackage{algorithm}
\usepackage{subfigure}
\usepackage{algorithmic}
\usepackage{epsfig}
\usepackage{float}
\usepackage{fullpage}

\begin{document}
\title{Bloomier Filters: A second look}
\author{Denis Charles \and Kumar Chellapilla}
\institute{Microsoft Live Labs, One Microsoft Way, Redmond WA - 98052. \\ \email{cdx@microsoft.com}, 
\email{kumarc@microsoft.com}
}
\date{26 March, 2008}

\maketitle
\begin{abstract}
A Bloom filter is a space efficient structure for storing static sets, where the space efficiency is gained at the expense of a small probability of false-positives. A {\em Bloomier filter} generalizes a Bloom filter to compactly store a function with a static support. In this article we give a simple construction of a Bloomier filter. The construction is linear in space
and requires constant time to evaluate. The creation of our Bloomier filter
takes linear time which is faster than the existing construction. We show how one can improve the space utilization further at the cost of increasing the time for creating the data structure. 
\end{abstract}
\section{Introduction}
A {\em Bloom filter} is a compact data structure that supports set membership queries \cite{bloom}.
Given a set $S \subseteq D$ where $D$ is a large set and $|S| = n$, the Bloom filter requires
space $O(n)$ and has the following properties. It can answer membership queries in $O(1)$ time.
However, it has one-sided error: Given $x \in S$, the Bloom filter will always declare that $x$ belongs to $S$, but given $x \in D \backslash S$ the Bloom filter will, with high probability,
 declare that $x \notin S$. Bloom filters have found wide ranging applications \cite{appl_1,appl_2,appl_3,appl_4,appl_5,appl_6}. There have also been generalizations in several 
 directions of the Bloom filter \cite{gen_1,gen_2,gen_3,gen_4}. 
 More recently, Bloom filters have been generalized to ``Bloomier'' filters that
 compactly store functions \cite{bloomier}.
In more detail: Given $S \subseteq D$ and a function $f: S \rightarrow \{0,1\}^k$
a Bloomier filter is a data structure that supports queries to the function value. It
also has one-sided error: given $x \in S$, it always outputs the correct value $f(x)$
and if $x \in D \backslash S$ with high probability it outputs `$\bot$', a symbol not
in the range of $f$. In \cite{bloomier} the authors construct a Bloomier filter
that requires, $O(n \log n)$ time to create; $O(n)$ space to store and, $O(1)$ time to evaluate.\\

In this paper we give an alternate construction of Bloomier filters,
which we believe is simpler than that of \cite{bloomier}. It has similar space
and query time complexity. 
The creation is slightly faster, $O(n)$ vs. $O(n \log n)$.
Changing the value of $f(x)$ while keeping $S$ the
same is slower in the worst case for our method, $O(\log n)$ vs. $O(1)$. 
For a detailed comparison we direct the reader to \S\ref{sec_comp:this}.
In \S\ref{sec_reduce:this} we discuss another construction that is very natural
and has a smaller space requirement. However, this algorithm has a creation time of $O(n^3)$ which
is too expensive. In \S\ref{sec_bucket:this} we discuss how bucketing can be used
to reduce the construction time of this algorithm to $n \log^{O(1)} n$ and make it more practical. 
In \S\ref{sec_experiment:this} we discuss
some experimental results comparing the existing construction to ours for
storing the in-degree information for a billion URLs.
\section{The construction}\label{sec_mainconstruct:this}
\subsection{A $1$-bit Bloomier Filter} We begin with the following simplified problem: 
Given a set $S$ of $n$ elements and
a function $f : S \rightarrow \{0,1\}$, encode $f$ into a space efficient
data structure that allows fast access to the values of $f$.
A simple way to solve this problem is to use a hash table (with open addressing) which requires $O(n)$ space and $O(1)$ time on average to evaluate $f$. If we want worst case $O(1)$ time for function evaluation, we could try different hash functions until we find one which produces few hash collisions on the set $S$. This solution however does not generalize to our ultimate goal which is to have a compact encoding of the function $\tilde{f} : D \rightarrow \{0,1,\bot\}$, where $\tilde{f}|_S = f $
 and $\tilde{f}(x) = \bot$ with high probability if $x \notin S$. Thus if $D$ is much larger than $S$, the solution using hash tables is not very attractive as it uses space proportional to $D$. To counter-act this one could use the hash table in conjunction with a Bloom filter for $S$. This is not the approach we will take\footnote{The reason this is not optimal is because to achieve error probability $\epsilon$, we will need
 to evalute $O(\log 1/\epsilon)$ hash functions.}.\\
 
Our approach to solving the simplified problem uses ideas from the creation of minimal perfect hashes (see \cite{chm92}). We first map $S$ onto the edges of a random (undirected) graph $G(V,E)$ constructed as follows. Let $V$ be a set of vertices with $|V| \geq c|S|$, where $c \geq 1$ is a constant. Let $h_1, h_2 : D \rightarrow V$ be two hash functions. For each $x \in S$, we create an edge $e = (h_1(x), h_2(x))$
and let $E$ be the set of edges formed in this way (so that $|E| = |S| =n$). If the graph $G$ is not acyclic we try again with two independent hash functions $h_1', h_2'$. It is known that if $c > 2$, then the
expected number of vertices on tree components is $|V| + O(1)$ (\cite{bol84} Theorem 5.7 ii). Indeed, in \cite{graph_acyclic} the authors proved that if $G(V,E)$ is a random graph
with $|V| = c |E|$ and $c > 2$, then with probability $\exp(1/c) \sqrt{(c-2)/c}$ the
graph is acyclic. Thus, if $c > 2$ is fixed then the expected number of iterations
till we find an acyclic graph is $O(1)$. In particular, if $c \geq 2.09$ then with probability at least $1/3$ the graph $G$ is acyclic.
Thus the expected number of times we will have to re-generate the graph until we find an acyclic graph is $\leq 3$. Once we have an acyclic graph $G$, we try to find a function $g : V \rightarrow \{0,1\}$ such that $f(x) \equiv g(h_1(x))+g(h_2(x)) \pmod 2$ for each $x \in S$. One can view this as a sequence of $n$ equations
for the variables $g(v)$, $v \in V$. The fact that $G$ is acyclic implies that
the set of equations can be solved by simple back-substitution in linear time. We then store the table of values $g(v)$ ($\in \{0,1\}$) for each $v \in V$. To evaluate the function $f$, given $x$, we compute $h_1(x)$ and $h_2(x)$ and add up the values stored in the table $g$ at these two indices modulo $2$. The expected creation time is $O(n)$, evaluation time is $O(1)$ (two hash function computations and two memory lookups to the table of values $g$) and the space utilization is $\lceil c n \rceil$ bits.\\

Next, we generalize this approach to encoding the function $\tilde{f} : D \rightarrow \{0,1,\bot\}$ that when restricted to $S$ agrees with $f$ and outside of $S$ it maps to $\bot$ with high probability. Here again we will use the same construction of the random acyclic graph $G(V,E)$ together with a map from $S \rightarrow E$ via two hash functions $h_1, h_2$. Let $m \geq 2$ be an integer and $h_3 : D \rightarrow \mathbb{Z}/m\mathbb{Z}$ be another independent hash function. We solve for a function $g : V \rightarrow \mathbb{Z}/m\mathbb{Z}$ such that the equations $f(x) \equiv g(h_1(x))+g(h_2(x)) + h_3(x) \pmod m$ holds for each $x \in S$. Again since the graph $G$ is acyclic these equations can be solved using back-substitution. Note that back-substitution works even though we are dealing with the ring $\mathbb{Z}/m\mathbb{Z}$ which is not a field unless $m$ is prime. To evaluate the function $f$ at $x$ we compute $h_i(x)$ for $1 \leq i \leq 3$ and then compute $g(h_1(x)) + g(h_2(x)) + h_3(x) \pmod m$. If the computed value is either $0$ or $1$ we output it otherwise, we output the symbol $\bot$. Algorithms \ref{alg_gentable:this} and \ref{alg_queryfunc:this} give the steps of the construction in more detail. It is clear that if $x \in S$ then the value output by our algorithm is the correct value $f(x)$. If $x \notin S$ then the value of $h_3(x)$ is independent of the values of $g(h_1(x))$ and $g(h_2(x))$ and uniform in the range $\mathbb{Z}/m\mathbb{Z}$. Thus $\Pr_{x \in D\backslash S}[g(h_1(x))+g(h_2(x))+h_3(x) \in \{0,1\}] = \frac{2}{m}$. 

\begin{algorithm}
\caption{Generate Table}
\label{alg_gentable:this}
\begin{algorithmic}
\REQUIRE A set $S \subseteq D$ and a function $f : S \rightarrow \{0,1\}$, $c > 2$, 
and an integer $m \geq 2$.
\ENSURE Table $g$ and hash functions $h_1,h_2,h_3$ such that $\forall s \in S: g[h_1(s)] + g[h_2(s)] + h_3(s) \equiv f(s) \mod m$.
\STATE Let $V = \{0,1,\cdots,\lceil c n\rceil -1 \}$, where $n = |S|$
\REPEAT
\STATE Generate $h_1, h_2 : D \rightarrow V$ where $h_i$ are chosen independently 
from $\mathcal{H}$ -- a family of hash functions; Let $E = \{(h_1(s),h_2(s)) ~:~ s \in S\}$.
\UNTIL{$G(V,E)$ is a simple acyclic graph.}
\STATE Let $h_3 : D \rightarrow \mathbb{Z}/m\mathbb{Z}$ be a third independently chosen hash function from $\mathcal{H}$.
\FORALL{$T$ -- a connected component of $G(V,E)$}
	\STATE Choose a vertex $v \in T$ whose degree is non-zero.
	\STATE $F \leftarrow \{v\}$; $g[v] \leftarrow 0.$ 
	\WHILE{$F \neq T$}
		\STATE Let $C$ be the set of nodes in $T \backslash F$ adjacent to nodes in $F$.
		\FORALL{$w = h_i(s) \in C$}
			\STATE $g[w] \leftarrow f(s) - g[h_{3-i}(s)] - h_3(s) \mod m$.
		\ENDFOR
		\STATE $F \leftarrow F \cup C$.
	\ENDWHILE
\ENDFOR
\end{algorithmic}
\end{algorithm}

In summary, we have proved the following:
\begin{proposition} \label{prp_basicbloomier:this}
Fix $c > 2$ and let $m \geq 2$ be an integer, 
the algorithms described above (Algorithms \ref{alg_gentable:this} and \ref{alg_queryfunc:this}) implement a Bloomier filter for storing the function $\tilde{f} : D \rightarrow \{0,1,\bot\}$ 
and the underlying function $f : S \rightarrow \{0,1\}$ with the following
properties:
\begin{enumerate}
\item The expected time for creation of the Bloomier filter is $O(n)$.
\item The space used is $\lceil c n \rceil \lceil \log_2 m\rceil$ bits, where $n = |S|$.
\item Computing the value of the Bloomier filter at $x \in D$ requires $O(1)$ time ($3$ hash function computations and $2$ memory lookups).
\item Given $x \in S$, it outputs the correct value of $f(x)$.
\item Given $x \notin S$, it outputs $\bot$ with probability $1 - \frac{2}{m}$.
\end{enumerate}
\end{proposition}
\subsection{General $k$-bit Bloomier Filters} It is easy to generalize the results of the previous section to obtain Bloomier filters
with range larger than just the set $\{0,1\}$. Given a function $f : S \rightarrow \{0,1\}^k$ it is clear that as long as the range $\{0,1\}^k$
embeds into the ring $\mathbb{Z}/m\mathbb{Z}$ one can still use Algorithm \ref{alg_gentable:this} without any changes. This translates into the simple requirement that we take $m \geq 2^k$. Algorithm \ref{alg_queryfunc:this} needs a minor modification, namely, we check if $f = g(h_1(x)) + g(h_2(x)) + h_3(x) \pmod{m} \in \{0,1\}^k$ and if so we output $f$ otherwise, we output $\bot$. We encapsulate the claims about the
generalization in the following theorem (the proof of which is similar to that of Proposition \ref{prp_basicbloomier:this}):
\begin{theorem} \label{thm_bloomier:this}
Fix $c > 2$ and let $m \geq 2^k$ be an integer, 
the algorithms described above implement a Bloomier filter for storing
the function $\tilde{f} : D \rightarrow \{0,1\}^k \cup \{\bot\}$, and the underlying function $f : S \rightarrow \{0,1\}^k$
with the following properties:
\begin{enumerate}
\item The expected time for creation of the Bloomier filter is $O(n)$.
\item The space used is $\lceil c n \rceil \lceil \log_2 m\rceil$ bits, where $n = |S|$.
\item Computing the value of the Bloomier filter at $x \in D$ requires $O(1)$ time ($3$ hash function computations and $2$ memory lookups).
\item Given $x \in S$, it outputs the correct value of $f(x)$.
\item Given $x \notin S$, it outputs $\bot$ with probability $1 - \frac{2^k}{m}$.
\end{enumerate}
\end{theorem}

\begin{algorithm}
\caption{Query function}
\label{alg_queryfunc:this}
\begin{algorithmic}
\REQUIRE Table $g$, $h_1, h_2 : D \rightarrow \{0, \cdots, \lceil cn\rceil-1\}$, $h_3 : D \rightarrow \mathbb{Z}/m\mathbb{Z}$ hash functions and $x \in D$.
\ENSURE $0, 1$ or $\bot$ -- the output of the Bloomier filter represented by the table $g$.
\STATE $f \leftarrow g[h_1(x)]+g[h_2(x)]+h_3(x) \mod m$.
\IF{$f \in \{0,1\}$}
	\STATE Output $f$.
\ELSE
	\STATE Output $\bot$.		
\ENDIF
\end{algorithmic}
\end{algorithm}
\subsection{Mutable Bloomier filters} In this section we consider the task of handling changes to the function stored in the Bloomier filter produced
by the algorithms in the previous section. We will only consider changes to the function $f : S \rightarrow \{0,1\}^k$ where $S$ remains the same
but only the values taken by the function changes. In other words, the support of the function remains static.\\

Consider what happens when $f : S \rightarrow \{0,1\}^k$ is changed to the function $f' : S \rightarrow \{0,1\}^k$ where $f(x) = f'(x)$ except
for a single $y \in S$. In this case we can change the values stored in the $g$-table so that we output the value of $f'$ at $y$. We assume that the edges of the graph $G$ are available (this is an additional $O(n\log n)$ bits). 
We begin with
the observation that the values stored at $g(v)$ for vertices $v$ not in the connected component containing the edge $e = (h_1(y),h_2(y))$ remain unaffected. 
Thus changing $f$ to $f'$ affects only the $g$ values of the connected component, $C$ (say), containing the edge $e$. Recomputing the $g$ values 
corresponding to $C$ would take time $O(|C|)$. How big can the largest connected component in $G$ get? Our graph $G(V,E)$ built
in Algorithm \ref{alg_gentable:this} is a sparse random graph with $|E| < \frac{1}{2}|V|$. A classical result due to Erd\H{o}s and R{\'e}nyi
says that in this case the largest component is almost surely\footnote{This means that the probability that the condition holds is $1-o(1)$.} $O(\log n)$ in size where $n = |E|$ (see \cite{er60} or \cite{bol84}). Thus updates
to the Bloomier filter take $O(\log n)$ time provided we ensure that the largest component in $G$ is small when creating it. The result from \cite{er60}
tells us that adding the extra condition while creating $G$ will not change the expected running time of Algorithm \ref{alg_gentable:this}. We call
this modified algorithm Algorithm \ref{alg_gentable:this}'.

\begin{theorem} The Bloomier filter constructed using algorithms \ref{alg_gentable:this}' and \ref{alg_queryfunc:this} can accomodate
changes to function values in time $O(\log n)$, provided the graph $G$ is also retained. Moreover, the claims of Theorem \ref{thm_bloomier:this} remain true
for algorithms \ref{alg_gentable:this}' and \ref{alg_queryfunc:this}.
\end{theorem}

\section{Reducing the space utilization}\label{sec_reduce:this}
If we are willing to spend more time in the creation phase of the Bloomier filter, we can
further reduce the space utilization of the Bloomier filter. 
In this section we show how one can get a Bloomier filter for a function $f : S \rightarrow \{0,1\}^k$ with error rate $\frac{2^k}{m}$ using only $n(1+\epsilon) \lceil \log_2 m \rceil$ bits of storage, where $n = |S|$ and $\epsilon > 0$ is a constant. 
In \S\ref{sec_mainconstruct:this} we used a random graph generated by 
hash functions to systematically generate a set of equations that can be solved efficiently.
The solution to these equations is then stored in a table which in turn encodes the function $f$.
The main idea to reduce space usage further is to have a table $g[0], g[1], \cdots, g[N-1]$, 
 where $N = (1+\epsilon)n$, and
try to solve the following set of equations over $\mathbb{Z}/m\mathbb{Z}$:
\begin{align}\label{eqn_setup:this}
	\left(\sum_{1 \leq i \leq s} h_i(x) g[h_{i+s}(x)] \right) + h_{0}(x) = f(x), ~x \in S
\end{align}
for the unknowns $g[0],\cdots, g[N-1]$. 
Here $s \geq 1$ is a fixed integer and
$h_0,h_1,\cdots, h_{2s}$ are independent hash functions. Since $s$ is fixed, look up of
a function value will only take $O(1)$ hash function evaluations. 
These equations
can be solved provided the determinant of the {\em sparse} matrix corresponding to
these equations is a unit in $\mathbb{Z}/m\mathbb{Z}$. The next subsection
gives an answer (under suitable conditions) to this question when $m$ is a prime.
\subsection{Full rank sparse matrices over a finite field} Let $\GL_{n\times r}^{s}(\mathbb{F}_p)$
be the {\em set} of full rank $n \times r$ matrices over $\mathbb{F}_p$\footnote{Here $p$ is
a prime number and $\mathbb{F}_p$ is
{\em the} finite field with $p$ elements. Any two finite fields with $p$ elements
are isomorphic and the isomorphism is canonical. If the field has $p^r$, $r>1$, elements then the isomorphism
is not canonical.} that have
exactly $s$ non-zero entries in each column. Our aim in this section is to get a lower bound for $\sharp \GL_{n\times r}^{s}(\mathbb{F}_p)$ (the cardinality of this set). We note the following lemma whose proof we omit.
\begin{lemma} Let $M_{n \times r}^{s}(\mathbb{F}_p)$ be the matrices over $\mathbb{F}_p$ where
each column has exactly $s$ non-zero entries. Then 
$
\sharp M_{n \times r}^{s}(\mathbb{F}_p) = \left(\binom{n}{s}(p-1)^s\right)^r.
$
\end{lemma}

Before we begin the task of getting a lower bound for the sparse full rank matrices we briefly recall the method of proof for finding $\sharp \GL_n(\mathbb{F}_p)$ -- the group of invertible
$n\times n$ matrices over $\mathbb{F}_p$. 
One can build invertible
matrices column by column as follows: Choose any non-zero vector for the first column, there are $p^n - 1$ ways of choosing the first column. The second column vector should not lie in the linear span
of the first. Therefore there are $p^n - p$ choices for the second column vector. Proceeding
in this way there are $p^n - p^j$ for the $j+1$ column. Thus we have
$
\sharp \GL_n(\mathbb{F}_p) = \prod_{1 \leq j \leq n}(p^n - p^{n-j}).
$\\

One can adapt this idea to get a bound on the invertible $s$-sparse matrices. There are $\binom{n}{s}(p-1)^s$ ways of choosing the first column. Inductively, suppose
we have chosen the first $i$ columns to be linearly independent, then we have
a vector space $V_i \subseteq \mathbb{F}_p^n$ of dimension $i$ spanned by the first $i$ columns.
One can grow this matrix to a rank $i+1$ matrix by augmenting it by any $s$-sparse vector $\mathbf{w} \notin V_i$. Thus we are faced with the task of finding an upper bound on the number
of $s$-sparse vectors contained in $V_i$. We introduce some notation: suppose 
$\mathbf{w} = \langle w_1, w_2, \cdots, w_n\rangle^t \in \mathbb{F}_p^n$ is a vector then we define $\mathbf{w}^{\oslash}$ to be the vector $\langle w_n, w_1, \cdots, w_{n-1} \rangle^t$ (a cyclic shift of $\mathbf{w}$). Note that if $\mathbf{w}$ is $s$-sparse then so is $\mathbf{w}^{\oslash}$. Our approach is to show that under certain circumstances the vector space spanned by the orbit of a sparse vector under the circular shifts have high dimension and consequently, all the shifts cannot be contained in $V_i$ (unless $i=n$). It is natural to expect that given a $s$-sparse vector $\mathbf{w}$, the vector space $W^{\oslash}$ spanned by all the circular shifts $\mathbf{w},\mathbf{w}^{\oslash},\cdots, \mathbf{w}^{\oslash^{n-1}}$ has dimension $\geq n-s$. Unfortunately, this is not so: For example, consider $\mathbf{w} = \langle 1, 0, 1, 0, 1, 0\rangle$ whose cyclic shifts generate a vector space of dimension $2$. This motivates the next lemma.

\begin{lemma} Suppose $q$ is a prime number and $\mathbf{w} \in \mathbb{F}_{p}^q$ is an $s$-sparse
vector with $0 < s < q$. Then the orbit $\{\mathbf{w},\mathbf{w}^{\oslash},\cdots,\mathbf{w}^{\oslash^{q-1}}\}$
has cardinality $q$.
\end{lemma}
\begin{proof} We have a natural action of the group $\mathbb{Z}/q\mathbb{Z}$
on the set of cyclic shifts of $\mathbf{w}$, via $a \mapsto \mathbf{w}^{\oslash^a}$. Suppose
we have $\mathbf{w}^{\oslash^i} = \mathbf{w}^{\oslash^j}$ for $0 \leq i \neq j \leq q-1$.
Then we have $\mathbf{w}^{\oslash^{(i-j)}} = \mathbf{w} = \mathbf{w}^{\oslash^q}$. Since
we have a group action this implies that $\mathbf{w}^{\oslash^{\gcd(i-j,p)}} = \mathbf{w}$.
Since $q$ is prime this means that $\mathbf{w}^{\oslash} = \mathbf{w}$. But $0 < s < q$
therefore $\mathbf{w}^{\oslash} \neq \mathbf{w}$ and we have a contradiction.
\qed \end{proof}


One can show that the vector space spanned by the cyclic shifts of an $s$-sparse vector ($0< s < n$) has dimension at least $n/s$. However, this bound is not sufficient for our purpose. We need the following stronger conditional result whose proof is relegated to the appendix (see Theorems \ref{thm_detnonvanish:this} and \ref{thm_critinert:this} in the Appendix).

\begin{theorem} Let $\mathbf{w} = \langle w_0,\cdots,w_{q-1} \rangle \in \mathbb{F}_p^q$, where $p$ is a prime that is a primitive root modulo $q$ (i.e., $p$ generates the cyclic group $\mathbb{F}_q^*$). 
Suppose $w_0 + w_1 \cdots + w_{q-1} \neq 0$ and $w_i$ are not all equal, then $W^{\oslash}$ (the vector space spanned by the cyclic shifts of $\mathbf{w}$) has dimension $q$.
\end{theorem}

Let $V_i$ be a vector space of dimension $i$ contained in $\mathbb{F}_p^q$. We have $\frac{1}{q}\binom{q}{s}(p-1)^s$ orbits of size $q$ under the action of $\mathbb{Z}/q\mathbb{Z}$ on the $s$-sparse vectors. 
If $s < n$ then all the coordinates cannot be identical. Once the $s$ non-zero positions for an $s$-sparse vector are chosen there are $\geq (p-1)^s - (p-1)^{s-1}$ vectors whose coordinates do not sum to zero\footnote{Indeed, it is not hard to show that the exact number of such vectors is $\frac{(p-1)\left((p-1)^s + (-1)^{s+1}\right)}{p}$.}. Now each of these orbits generates a vector space of rank $q$
by the above theorem. In each orbit there are at most $i$ vectors that can belong to $V_i$. Consequently, there are at least
\begin{align}\label{eqn_rankinc:this}
\frac{1}{q}\binom{q}{s}\left((p-1)^s - (p-1)^{s-1} \right)(q-i)
\end{align}
$s$-sparse vectors that do not belong to $V_i$. We have thus proved the following theorem:

\begin{theorem} \label{thm_sparse_lowbound:this}
Let $q, p$ be prime numbers such that $p$ is congruent to a primitive root modulo $q$.
Then
\begin{align*}
	\sharp \GL_{q\times r}^{s}(\mathbb{F}_p) \geq 
	\prod_{0 \leq i \leq r-1} \left( \frac{1}{q}\binom{q}{s}\left((p-1)^s - (p-1)^{s-1}\right)(q-i) \right).
\end{align*}
\end{theorem}

We note that the bound obtained above is almost tight\footnote{The bound is tight if we use the exact formula for the number of $s$-sparse vectors that do not sum to $0$ in the derivation.} in the case $s = 1$, where the $1$-sparse matrices
are simply diagonal matrices (with non-zero entries) multiplied by permutation matrices.
\subsection{The Algorithm}
The outline of the algorithm is as follows. To create the
Bloomier filter given $f : S \rightarrow \{0,1\}^k$, we consider each element $x$ of $S$ in
turn. We generate a random equation
as in (\ref{eqn_setup:this}) for $x$ and check that the list of equations that we have
so far has full rank. If not, we generate another equation using a different set
of $2s$ hash functions. At any time, we keep the hash functions that have been used
so far in blocks of $2s$ hash functions. When generating a new equation we always
start with the first block of hash functions and try subsequent blocks only if the previous blocks
failed to give a full rank system of equations. The results of the previous section
show that the expected number of blocks of hash functions is bounded (provided the vector space has high dimension). Once we have
a full rank set of equations for all the elements of $S$, we then proceed to solve
the sparse set of equations. The solution to the equations is then stored in a table.
At look up time, we generate the equations using each block of hash functions in turn and
output the first time an equation generates a value in the range of $f$.
At first glance it looks like this approach stores $f$ with two-sided error, i.e., even
when given $x \in S$ we might output a wrong value for $f(x)$. However, we
show that the probability of error committed by the procedure 
on elements of $S$ can be made so small
that, by doing a small number of trials, we can ensure that we do not err on any element of $S$.

\begin{algorithm}
\caption{Setup parameters}
\label{alg_setupparams:this}
\begin{algorithmic}
\REQUIRE $n \geq 0$ integer given in unary, $m \geq 0$ integer, $\epsilon > 0$.
\ENSURE $q$ and $p$ primes, $p$ is a $m$-bit prime that is a primitive root modulo $q$.
\STATE Let $q$ be the first prime $\geq n(1+\epsilon)$.
\STATE Factor $q-1$ and let $q_1,\cdots,q_k$ be the (distinct) prime factors of $q-1$.
\REPEAT
	\STATE Choose a random $g \in \mathbb{F}_q$.
\UNTIL{$g^{q_i} \not\equiv 1 \mod q$ for each $1 \leq i \leq k$.}
\STATE Let $g_i = g^{i} \mod q$ for $1 \leq i \leq q-1$, $\gcd(i,q-1)=1$.
\REPEAT
	\STATE Choose a random $m$-bit integer $p$.
\UNTIL{$p \equiv g_i$ for some $i$, and $p$ is prime.}
\end{algorithmic}
\end{algorithm}

{\bf Analysis of Algorithm \ref{alg_setupparams:this}:} The first step
of the algorithm finds the smallest prime larger than $n(1+\epsilon)$. The prime
number theorem implies that the average gap between primes is $\ln n$, this
means that on average $q \leq n(1+\epsilon)+\ln n$. Let $\pi(x)$
be the prime counting function. Then from the prime number theorem 
$\pi(x+\delta x) - \pi(x) \sim \frac{\delta x}{\ln x}$, for any fixed $\delta > 0$ (much
stronger results are known, see \cite{bhp01}). Thus, if $n$ is large enough $q \leq (1+\epsilon')n$
for any $\epsilon' > \epsilon$. Since $n$ is provided in unary, the algorithm can
factor $q-1$ in linear time to obtain the prime factors $q_1, \cdots, q_k$. The loop following
this step finds a primitive root modulo $q$. Since $\mathbb{F}_{q}^*$ is cyclic of order $q-1$,
standard group theory tells us that there are $\varphi(q-1)$ such generators, where $\varphi$
is Euler's totient function. It is known that $\varphi(n) \gg \frac{n}{\log \log n}$ (see \cite{rs62})
and so the expected number of times the loop runs is $O(\log \log q)$. Once a generator
$g$ is found, the algorithm computes the list $g_i$ that are all generators of the group $\mathbb{F}_q^*$. This step requires $O(n \log^2 n)$ as we do $O(n)$ arithmetic
operations over the field $\mathbb{F}_q$.
The final loop of the algorithm attempts to find a random prime $p \equiv g_i$. By
the prime number theorem for arithmetic progressions the number of primes
that are $\equiv g_i \mod q$, for some $i$, below a bound $x$ is aymptotic to $\frac{\varphi(q-1)x}{(q-1)\ln x}$. Thus a random number in the interval $[1 \cdots x]$ satisfies the termination condition for the loop with probability $\frac{\varphi(q-1)}{(q-1) \ln x}$.
However, we seek a prime of $m$-bits and so we pick numbers at random in the interval 
$[2^{m-1} \cdots 2^m-1]$. Again, by the prime number thorem for arithmetic progressions,
the number of primes in this interval is $\sim \frac{\varphi(q-1) x}{2(q-1)(\ln x - \ln 2)}$,
where $x = 2^m -1$. This tells us that the expected number of iterations of the loop is 
about $\frac{2\ln 2 (q-1)(m-1)}{\varphi(q-1)} = O(m \log \log q)$. We can use a probabilistic
primality test to check for primality of the random $m$-bit numbers that
we generate. If we use the Miller-Rabin primality test (from \cite{rab80}) the expected
number of bit-operations\footnote{The soft-Oh notation, $\tilde{O}$,
hides factors of the form $\log\log n$ and $\log m$} is $\tilde{O}(m^4)$. In summary, the expected running time of Algorithm \ref{alg_setupparams:this} is $\tilde{O}(n + m^4)$.
Note that $m$ will be very small in practice (a prime of about $64$-bits should suffice).
\begin{algorithm}
\caption{Create Table}
\label{alg_createtable:this}
\begin{algorithmic}
\REQUIRE A set $S \subseteq D$ and a function $f : S \rightarrow \{0,1\}^k$, two primes $p, ~q$, $\mathcal{H}$ a hash family, and $s \geq 2$.
\ENSURE Table $\mathbf{g}$, $h_0$ a hash function and $r$ blocks of $2s$ hash functions $B_i$.
\STATE $M \leftarrow (0)_{n \times q}$ (a $n \times q$ zero matrix). 
\STATE Let $h_0$ be a random hash function from $\mathcal{H}$.
\STATE $i \leftarrow 0.$
\FORALL{$x \in S$}
	\STATE $i \leftarrow i+1$; $j \leftarrow 0$.
	\REPEAT
		\IF{$B_j$ is not defined}
			\STATE Generate $h_1,\cdots, h_{2s}$ random hash functions from $\mathcal{H}$.
			\STATE $B_j \leftarrow \{h_1,\cdots, h_{2s} \}$.
		\ENDIF
		\STATE Let $h_1,\cdots,h_{2s}$ be the hash functions in $B_j$.
		\STATE $M[i,h_{k+s}(x)] \leftarrow h_k(x)$ for $1 \leq k \leq s$; $j \leftarrow j+1$.
	\UNTIL{$\mathrm{Rank}(M) = i$}
\ENDFOR
\STATE Let $\mathbf{v} = \langle f(x) - h_0(x) : x \in S\rangle^t$.
\STATE Solve the system $M \times \mathbf{g} = \mathbf{v}$ for $\mathbf{g} = \langle g[i] : 1 \leq i \leq q\rangle^t$ over $\mathbb{F}_p$.
\STATE {\bf Return} $\mathbf{g}$, $h_0$ and $B_i$.
\end{algorithmic}
\end{algorithm}

{\bf Analysis of Algorithm \ref{alg_createtable:this}:} The algorithm essentially mimics
the proof of Theorem \ref{thm_sparse_lowbound:this}. It starts with a rank $i$ matrix
and grows the matrix to a rank $i+1$ matrix by adding an $s$-sparse row using hash functions in $B_j$\footnote{Strictly speaking the row could have $< s$ non-zero entries because a hash function could map to zero. But this happens with low probability.}. Let $n = |S|$ and suppose, $q \geq n(1+\epsilon)$ for a fixed $\epsilon > 0$. Then equation (\ref{eqn_rankinc:this}) tells us that in $O(1/\epsilon)$ iterations we will find that
the rank of the matrix increases. In more detail, the probability that a random $s$-sparse
vector does not lie in $V_i$ is at least $\frac{q-i}{q} \geq \epsilon$ since $i < n$ and $q \geq n(1+\epsilon)$. 
Note that this requires rather strong pseudorandom
properties from the hash family $\mathcal{H}$. As mentioned in the discussion
following Lemma 4.2 in \cite{bloomier}, a family of cryptographically strong hash functions
is needed to ensure that the vectors generated by the hash function from the input
behave as random and independent sparse vectors over the finite field. We will make this
assumption on the hash family $\mathcal{H}$. Checking the rank can be done by Gaussian
elimination keeping the resulting matrix at each stage. The inner-loop thus runs in
expected  $O(n^2)$ time and the ``for'' loop takes $O(n^3)$ time on average. Solving
the resulting set of sparse equations can be done in $O(n^2)$  time since the Gaussian
elimination has already been completed. The algorithm also generates $r$ blocks of hash functions,
and by the earlier analysis the expected value of $r$ is $O(1/\epsilon)$. In summary, the expected
running time of Algorithm \ref{alg_createtable:this} is $O(n^3)$. We refer the reader to the appendix for a discussion
on why sparse matrix algorithms cannot be used in this stage, and also why $s = 1$ cannot be used here.

{\bf Analysis of Algorithm \ref{alg_query:this}:} In this algorithm we try the blocks
of hash functions and output the first ``plausible'' value of the function (namely, a value
in the range of the function $f$). If the wrong block, $B_i$, of hash functions was used then
the probability that the resulting function value, $y$, belongs to the range $\{0,1\}^k$
is $\frac{2^k}{p}$. If the right block $B_i$ was used then, of course, we get the
correct value of the function and $y = f(x)$. If $x \in D \backslash S$, then again
the probability that $y \in \{0,1\}^k$ is at most $\frac{r 2^k}{p}$. Since $r$ and $s$
are $O(1)$ the algorithm requires $O(1)$ operations over the finite field $\mathbb{F}_p$. This
requires $O(\log^2 p)$ bit operations with the usual algorithms for finite field operations, and only  $O(\log p \log \log p)$ bit operations if FFT multiplication is used. 
\begin{algorithm}
\caption{Query function}
\label{alg_query:this}
\begin{algorithmic}
\REQUIRE Table $g$, hash functions $h_0, B_i, 1 \leq i \leq r$, $x \in D$.
\ENSURE $y \in \{0,1\}^k$ or $\bot$.
\STATE $i \leftarrow 1$
\WHILE{$i \leq r$}
	\STATE Let $h_1,\cdots,h_{2s}$ be the hash functions in $B_i$.
	\STATE Let $y \leftarrow h_0(x) + \sum_{1 \leq j \leq s} h_i(x) g[h_{i+s}(x)]$.
	\IF{$y \in \{0,1\}^k$}
		\STATE {\bf Return} $y$.
	\ENDIF
	\STATE $i \leftarrow i+1$
\ENDWHILE
\STATE {\bf Return} $\bot$.
\end{algorithmic}
\end{algorithm}

{\bf How to get one-sided error: } The analysis in the previous paragraph
shows that the probability that we err on any element of $S$ is $\leq \frac{n 2^k}{p}$.
Thus, if $p$ is large we can construct a $g$ table using Algorithm \ref{alg_createtable:this}
and verify whether we give the correct value of $f$ for all elements of $S$. If not,
we can use Algorithm \ref{alg_createtable:this} again to construct another table $g$. 
The probability we succeed
at any stage is $\geq 1 - \frac{n 2^k}{p}$, and if $p$ is taken large enough that this is $\geq \frac{1}{2}$, then the expected number of iterations is $\leq 2$.
We summarize the properties of the Bloomier filter constructed in this section below:
\begin{theorem}\label{thm_reduce_one_sided:this} Fix $\epsilon > 0$ and $s \geq 2$ an integer, 
let $S \subseteq D$, $|S| = n$ and let $m, k$ be positive integers
such that $m \geq k$.
Given $f : S \rightarrow \{0,1\}^k$, 
the Bloomier filter constructed, (with parameters $\epsilon, m$ and $s$) by Algorithms
\ref{alg_setupparams:this} and \ref{alg_createtable:this}, 
and queried, using Algorithm \ref{alg_query:this}, 
has the following properties:
\begin{enumerate}
\item The expected time to create the Bloomier filter is $\tilde{O}(n^3 + m^4)$.
\item The space utilized is $\lceil n (1+\epsilon) \rceil m$ bits.
\item Computing the value of the Bloomier filter at $x \in D$ requires $O(1)$ hash function
evaluations and $O(1)$ memory look ups. 
\item If $x \in S$, it outputs the correct value of $f(x)$.
\item If $x \notin S$, it outputs $\bot$ with probability $1 - O(\frac{1}{\epsilon}2^{k-m})$.
\end{enumerate}
\end{theorem}
\section{Bucketing}\label{sec_bucket:this}
The construction in \S\ref{sec_reduce:this} is space efficient but the time to construct the
Bloomier filter is exhorbitant. In this section we show how to mitigate this with bucketing. 
To build a Bloomier filter for $f: S \rightarrow \{0,1\}^k$,
one can choose a hash function $g : S \rightarrow \{0,1,\cdots, b-1\}$ and
then build Bloomier filters for the functions $f_i : S_i \rightarrow \{0,1\}^k$, for $i = 0,1,\cdots, b-1$, where $S_i = g^{-1}(i)$ and $f_i(x) = f(x)$ for $x \in S_i$. The sets $|S_i|$ have an expected size of $|S|/b$ and hence results in a speedup for the construction time. The bucketing also allows one to parallelize of the construction process, since each of the buckets can in processed independently. To quantify the time saved by bucketing we need a concentration result for the size of the buckets produced by the hash function.\\

Fix a bucket $b_i$, $0 \leq b_i < b$ and define random variables $X^{(b_i)}_{s_1}, \cdots, X^{(b_i)}_{s_n}$ for $s_j \in S$ as follows: Pick a hash function $g : S \rightarrow \{0,1,\cdots, b-1\}$ from a family of hash functions $\mathcal{H}$
and set $X^{(b_i)}_{s_j} = 1$ if $g(s_j) = b_i$ and set $X^{(b_i)}_{s_j} = 0$ otherwise.
Under the assumption that the random variables $X^{(b_i)}_{s_j}$ are mutually independent, we obtain
using Chernoff bounds that 
$
\Pr\left[\sum_{j} X^{(b_i)}_{s_j} > (1+\delta) \frac{|S|}{b}\right] < 2^{- \frac{\delta|S|}{b} }
$
provided $\delta > 2e - 1$. This bound holds for any bucket and consequently,
$
\Pr\left[\exists j~:~ \sum_{j} X^{(b_i)}_{s_j} > (1+\delta) \frac{|S|}{b}\right] < b2^{-\frac{\delta|S|}{b}}.
$
Thus with probability $> 1 - b2^{-\frac{\delta|S|}{b}}$ all the buckets have at most $(1+\delta) \frac{|S|}{b}$ elements. Suppose we take the number of buckets $b$ to be $\frac{|S|}{c \log |S|}$ for $c > 1$. Then the probability that all the buckets are of size $< c(1+\delta) \log |S|$ 
is at least $1 - 2^{(-c\delta \log |S| + \log |S| - c\log \log |S|)}$ which for large enough $S$ is
$> 1/2$. In other words, the expected number of trials until we find a hash function $g$ that results in all
the buckets being ``small'' is less than $2$. 

\begin{remark} Note that the assumption that the random variables $X_{s_j}$
be mutually independent requires the hash family to have strong pseudorandom properties. For instance, if $\mathcal{H}$ is a $2$-universal family of hash functions then the random variables $X_{s_j}$ are only pairwise independent. 
\end{remark}

In the following discussion we adopt the notation from Theorem \ref{thm_reduce_one_sided:this}. 
We assume that we have a hash function $g$ that results in all buckets have $O(\log n)$ elements. The time for creation of the Bloomier filter in \S\ref{sec_reduce:this} for each bucket is reduced to $O(\log^3 n + r^4)$. To query the bucketed Bloomier filter, given $x$, we first
compute the bucket, $g(x)$, and then query the Bloomier filter for that bucket.
Thus, querying requires one more hash function evaluation than the non-bucketing version.
 Suppose $n_i$
is the number of elements of $S$ that belonged to the bucket defined by $b_i$, then the
Bloomier filter for this bucket requires $\lceil n_i (1+\epsilon)\rceil r$ bits. The total
number of bits used is $\sum_{0 \leq i < b}\lceil n_i (1+\epsilon)\rceil r \leq \sum_{0 \leq i < b} \left(n_i(1+\epsilon) + 1 \right)r = n(1+\epsilon) r + b r$, since $\sum_i n_i = n$. Since the
number of buckets is $O(n/\log n)$, the number of bits used is $n(1+\epsilon)r + O(r n/\log n)$.\\

We summarize the properties of the bucketing variant of the construction in \S\ref{sec_reduce:this} in the
following theorem.

\begin{theorem}\label{thm_bucketed_one_sided:this} Fix $\epsilon > 0$ and $s \geq 2$ an integer, 
let $S \subseteq D$, $|S| = n$ and let $m, k$ be positive integers
such that $m \geq k$.
Given $f : S \rightarrow \{0,1\}^k$, bucketed using $|S|/c\log|S|$ buckets for a fixed $c>1$,
the Bloomier filter constructed on the buckets, (with parameters $\epsilon, m$ and $s$) by Algorithms
\ref{alg_setupparams:this} and \ref{alg_createtable:this}, 
and queried (on the buckets), using Algorithm \ref{alg_query:this}, 
has the following properties:
\begin{enumerate}
\item The expected time to create the Bloomier filter is $\tilde{O}(\frac{n}{\log n} (\log^3 n + m^4))$.
\item The space utilized is $n(1+\epsilon)m + O(m n/\log n)$ bits.
\item Computing the value of the Bloomier filter at $x \in D$ requires $O(1)$ hash function
evaluations and $O(1)$ memory look ups. 
\item If $x \in S$, it outputs the correct value of $f(x)$.
\item If $x \notin S$, it outputs $\bot$ with probability $1 - O(\frac{1}{\epsilon}2^{k-m})$.
\end{enumerate}
\end{theorem}

\begin{figure}
\centering
\subfigure[Creation Time for the Bloomier Filter]{
\epsfig{file=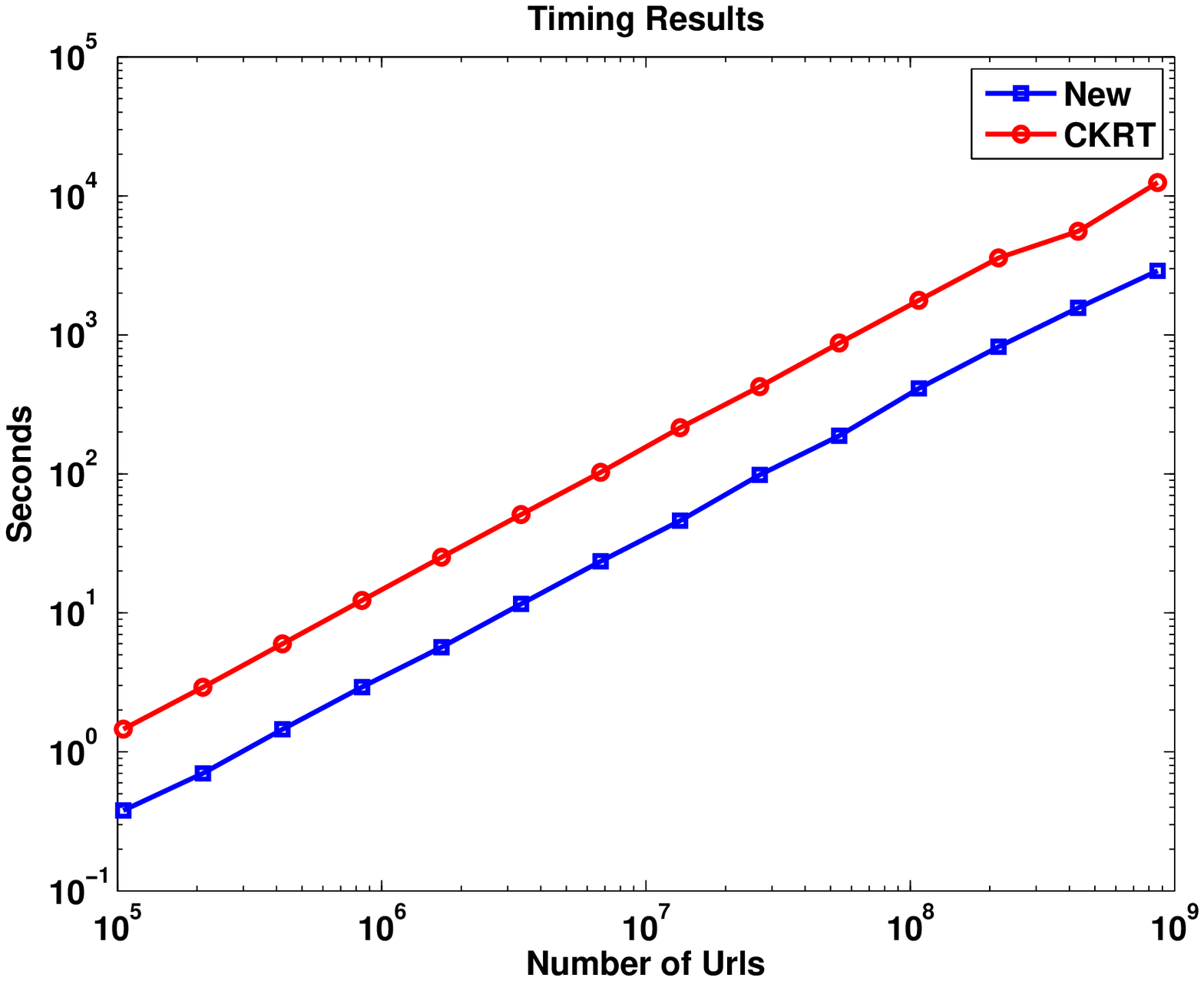, width=5cm}
}
\subfigure[The number of retries in the creation stage]{
\epsfig{file=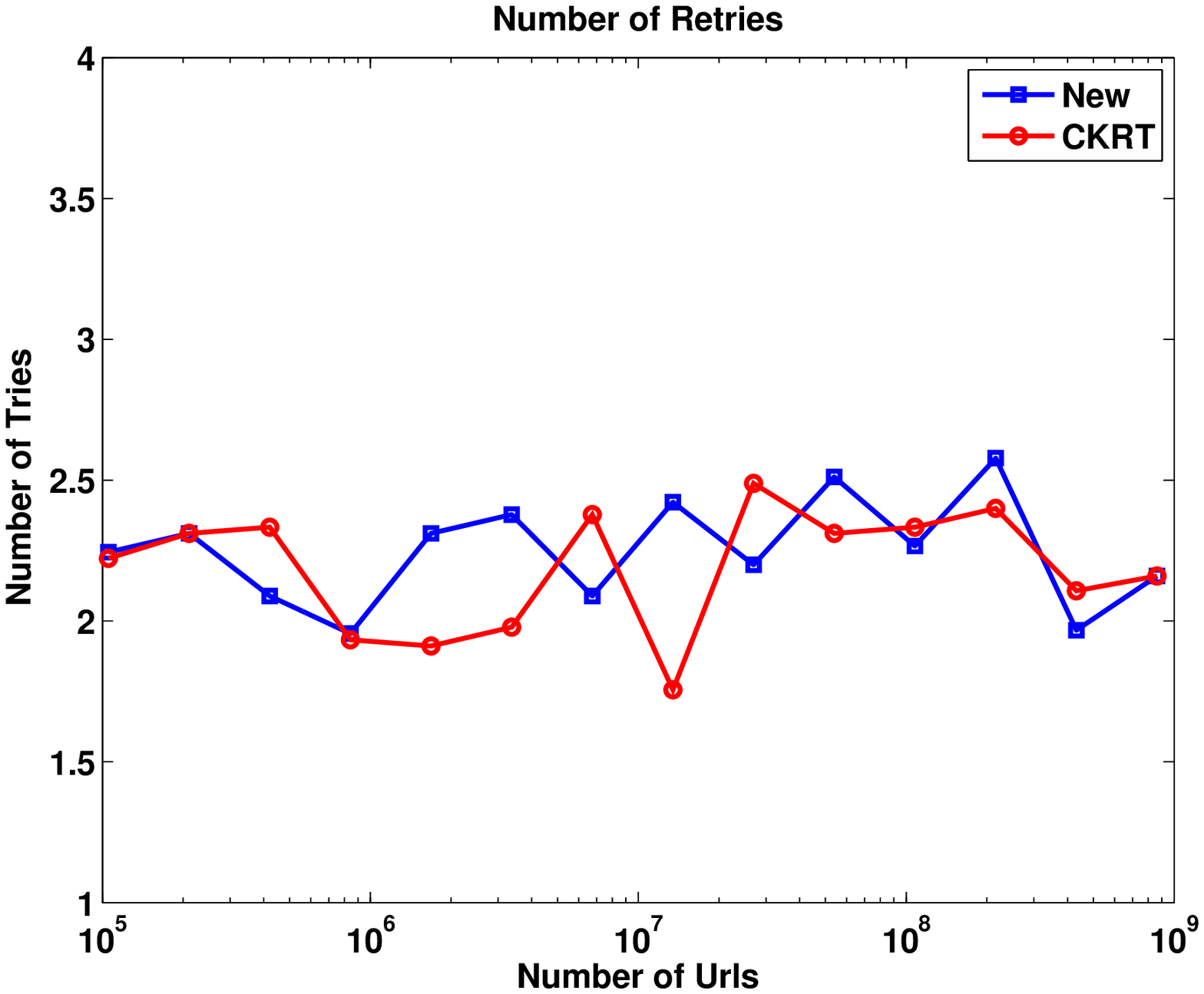, width=5cm}
}
\subfigure[The Speedup over the existing scheme]
{
\epsfig{file=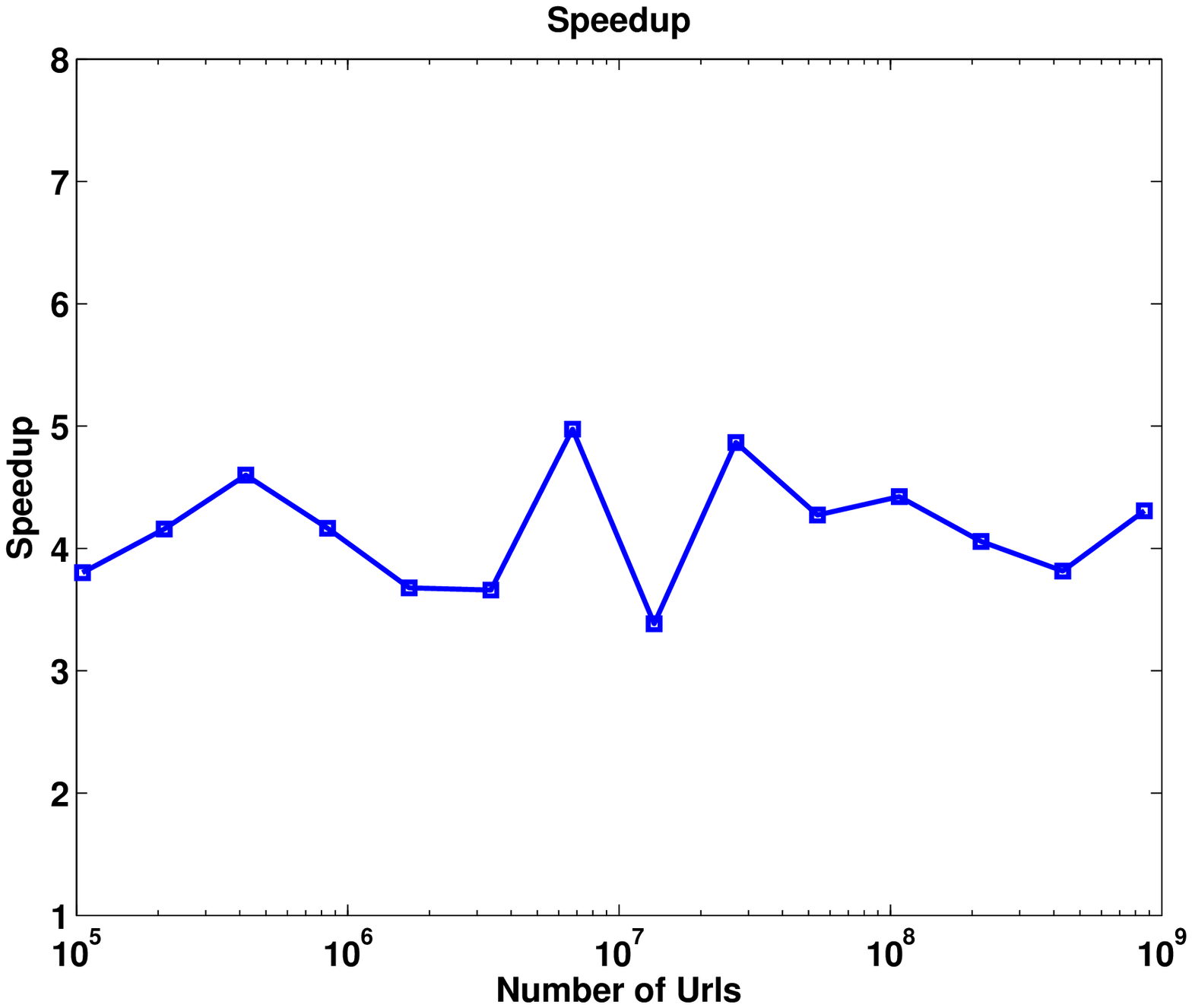, width=5cm}
}
\caption{Comparisons of the Bloomier Filters}
\label{fig_comparison:this}
\end{figure}

\section{A remark on the use of sparse matrices}
There are more efficient algorithms for computing the rank of a sparse
matrix and for inverting such matrices (see \cite{wie86,kal95,lo90}). However, these
algorithms do not lend themselves to an incremental operation: Fix $s$, given an $s$-sparse $r \times n$ matrix $M$ whose rank has already been computed by any of these algorithms, compute the rank of an $s$-sparse $(r+1) \times n$ matrix that contains $M$ in the first $r$ rows. To reduce
the running time of Algorithm \ref{alg_createtable:this} to $O(n^2)$, one would need
to solve the above problem in $O(n)$ time. Checking that
an $s$-sparse $n\times n$ matrix over $\mathbb{F}_p$ is full rank can be done in $O(n^2)$ time.
However, since the $s$-sparse invertible matrices are relatively rare (consider the case when $s$ is small, for instance, $1$) we cannot simply pick a random $s$-sparse matrix and have a resonable
probability of it being invertible. This is why we have to build the matrix in stages as we
do in the above algorithm.\\

One may wonder if the construction works for $s = 1$, because in this case the matrix inversion is easy. However, the
analysis of Algorithm 4 fails in this case. In more detail, if $s = 1$ then each block of hash functions contains two hash functions. Out of these, only one of them determines the ``variable'' $g[i]$ to be used for an element. The construction then attempts to implicitly create a matching between the $n$ elements of $S$ to these $g[i]$ of which there are $n(1+\epsilon)$ using $r$ hash functions.
If the number of hash function blocks $r$ were $O(1)$, then in particular there is a hash function $h_j$ that creates a matching for $n/r$ elements. This implies that a hash function is able to create an injection of a set $S'$ with $n/r$ elements into a set of size $n(1+\epsilon)$. However, the probability that this event occurs is exponentially low if $r = o(\sqrt{n})$, since by the Birthday paradox the expected number of colliding pairs in such a construction is $\binom{n/r}{2}/n(1+\epsilon)$.

\section{Comparison with earlier work}\label{sec_comp:this}
Our constructions in \S\ref{sec_mainconstruct:this} and \S\ref{sec_reduce:this}
and those of \cite{bloomier} are related in the broad sense that all approaches
use hash functions to generate equations that can be solved to construct a Bloomier filter.
However, the details differ markedly. Let $S \subseteq D$ and $f : S \rightarrow \{0,1\}^k$ be the function that we wish to store.
The approach of \cite{bloomier} is as follows: They first show that if we have $r$
hash functions (we need $r > 1$ see the discussion below), 
then for each element $x \in S$ we can single out
a hash value ($h_{i(x)}(x)$) which does not collide with the chosen hash values for the other elements. They
prepare a table that stores the mapping from $x$ to the chosen hash function ($i \mapsto i(x)$) efficiently,
and then look up a second table using the hash function indicated by the first table. A bit-mask
computed using another hash function is used to provide error resiliency. 
To find the ``chosen'' hash value for each element requires a matching problem to be solved.
For the matching problem to have a solution we need at least two hash functions. Indeed, if
we had only one hash function then the Birthday paradox implies that there will be collisions
among the elements (since the hash functions map the set $S$ of size $n$ to a set of size $\approx n$). For the colliding elements we cannot select the ``chosen'' hash value. 
Provided $r > 1$, they show that the matching problem can be solved 
in $O(n \log n)$ time on average. The space
used is $r c (q + k) n$ bits, where $r \geq 2$, $c > 1 + \frac{1}{\sqrt{n}}$ and $q = \log\frac{r}{\epsilon}$ (here $\epsilon$ is the probability of an error given $x \in D \backslash S$).
Look up requires $r+1$ hash function evaluations and $r+1$ memory accesses ($r$ accesses to $Table_1$
and one access to $Table_2$ in the notation of \cite{bloomier}). Since $r \geq 2$, we need
at least $3$ memory accesses. More importantly,
their construction allows changes to the function value in the same time as a look up.\\

The method from \S\ref{sec_mainconstruct:this} can be constructed in {\em linear} time
on average which is faster than \cite{bloomier}. The space utilization is similar $\approx (2+\delta) n(\log \frac{1}{\epsilon} + k)$ (for any $\delta > 0$). Changing the value of $f(x)$ for $x \in S$ is slower taking $O(\log n)$ time. Looking up a function value requires $3$ hash evaluations and $2$ memory accesses (which
is slightly faster than their scheme). 
The method from \S\ref{sec_reduce:this} is more efficient in the storage space than
both methods. It requires only $(1+\delta) n (\log \frac{1}{\epsilon} + k)$ space for any fixed $\delta > 0$. Look up time is still $O(1)$, but creation time is an exhorbitant $O(n^3)$. The bucketing
approach from \S\ref{sec_bucket:this} reduces this to $n \log^{O(1)} n$. 
We believe it would be an interesting
problem to construct Bloomier filters that require $c n (\log \frac{1}{\epsilon} + k)$ bits of storage for $c < 2$, while allowing a look up time of $O(1)$ and a creation time of $O(n)$.\\

{\bf Note: } Some recent independent results of \cite{pagh} are much closer to our approach. They use results of \cite{calkin} to get a bound on the number sparse invertible matrices over the field $\mathbb{F}_2$ and, consequently, encode the function.

\section{Experimental Results}\label{sec_experiment:this}
In this section we discuss the results of some experiments that we ran comparing our construction
of Bloomier filters (from \S\ref{sec_mainconstruct:this} ) and the scheme of \cite{bloomier}. The function we store is $InDeg$ that
maps a URL to the number of URLs that link to it. We obtained the in-link information for little over a billion URLs from the Live Search crawl data. 
We measured the creation time and memory usage for
both the schemes for various numbers of URLs and averaged the results over $45$ trials. 
The results are graphed in Figure \ref{fig_comparison:this}. 
Figure \ref{fig_comparison:this}a
shows the time taken by both methods for creation of the Bloomier filter (we used
$c = 2.5$ and error probability $= 2^{-32}$ for both the schemes). Figure \ref{fig_comparison:this}b
displays the number of trials by the creation phase of each algorithm to find an appropriate graph (acyclic in our algorithm and lossless expander in theirs). As one can see from the results, the number
of trials until a lossless expander is found is about the same as that of finding an acylic graph. However, it takes comparatively longer to find a matching in the graph. Our scheme ends up being between $3$ to $5$ times
faster for creation of the filter as a result (see \ref{fig_comparison:this}c). Also, the memory foot print of the constructed Bloomier filter in both schemes is similar allowing the in-link information for $\approx 1.1 \times 10^9$ URLs to
fit in $\approx 20$GB.

\appendix
\section[A]{Circulant matrices over finite fields}
\begin{definition} Let $V$ be an $n$-dimensional vector space over a field $\mathbb{F}$
and let $\mathbf{w} = \langle w_0, w_1, \cdots, w_{n-1} \rangle \in V$. A {\em circulant
matrix} associated to $\mathbf{w}$ is the $n \times n$ matrix
\begin{align*}
\begin{pmatrix}
w_0 & w_1 & \cdots & w_{n-2} & w_{n-1} \\
w_{n-1} & w_{0} & \cdots & w_{n-3} & w_{n-2} \\
\vdots & \vdots & \ddots & \vdots & \vdots \\
w_2 & w_3 & \cdots & w_0 & w_1 \\
w_1 & w_2 & \cdots & w_{n-1} & w_0
\end{pmatrix}
\end{align*}
\end{definition}

The following results are from \cite{circulant}, however, the proofs need some modification
since we are dealing with vector spaces over finite fields.\\

First we need a closed form for the determinant of a circulant matrix.

\begin{theorem}\label{thm_detcirculant:this}
Let $p$ be a prime and $n$ a positive integer relatively prime to $n$. 
Let $\mathbf{w} = \langle w_0, w_1, \cdots, w_{n-1} \rangle$ be a vector in $\mathbb{F}_p^n$
and let $W$ be the circulant matrix associated to $\mathbf{w}$. Then
\begin{align*}
\det W = \prod_{0 \leq \ell \leq n-1 } \left( \sum_{0 \leq j \leq n-1} \epsilon^{j \ell} w_j\right),
\end{align*}
where $\epsilon$ is a primitive $n$-root of unity contained in the algebraic closure $\overline{\mathbb{F}}_p$.
\end{theorem}
\begin{proof}
One can view $W$ as a linear transformation acting on the vector space $\overline{\mathbb{F}}_p^n$.
An explicit calculation shows us that the vectors $\mathbf{x}_{\ell} = \langle 1,
 \epsilon^{\ell}, \epsilon^{2\ell}, \cdots, \epsilon^{(n-1)\ell}\rangle$, $0 \leq \ell \leq n-1$, 
  are all eigenvectors with eigenvalues 
\begin{align*}
\lambda_{\ell} = w_0 + \epsilon^{\ell} w_1 + \cdots + \epsilon^{(n-1)\ell} w_{n-1}.
\end{align*}
Since $\{\mathbf{x}_0,\cdots,\mathbf{x}_{n-1}\}$ is a linearly independent set, we conclude
that
\begin{align*}
\det W = \prod_{0 \leq \ell \leq n-1} \lambda_{\ell}.
\end{align*}
It is not immediately apparent that the product is actually in $\mathbb{F}_p$. To show that
one looks at the smallest field $\mathbb{F}_{p^r}$ that contains $\epsilon$. The Galois
group of this field $\Gal(\mathbb{F}_{p^r}/\mathbb{F}_p)$ is cyclic of order $r$, indeed,
$r$ is the smallest integer such that $p^r \equiv 1 \mod n$. 
The Galois group is generated by the Frobenius map, $\mathrm{Fr}$, that sends $x \mapsto x^p$. Under this map, $\epsilon \mapsto \epsilon^p = \epsilon^{s}$ where $s \equiv p \mod n$. Consequently,  $\mathrm{Fr}(\lambda_{\ell}) = \lambda_{\ell'}$ where $\ell' \equiv p\ell \mod n$.
Since $\gcd(n,p) = 1$ the Frobenius just permutes the terms of the product $\prod_{0 \leq n \leq n-1} \lambda_{\ell}$.
The product $\prod_{0 \leq \ell \leq n-1} \lambda_{\ell} = \det W$  is fixed by the Galois group and so it belongs to $\mathbb{F}_p$.
\qed \end{proof}

\begin{theorem} \label{thm_detnonvanish:this}
Let $p$, $q$ be primes such that $q$-th cyclotomic
polynomial $f(x) = \sum_{0 \leq i < q} x^i$
is irreducible modulo $p$. Suppose $W$ is a circulant matrix associated
to $\mathbf{w} =\langle w_0, w_1,\cdots, w_{q-1}\rangle$ with entries
in the finite field $\mathbb{F}_p$, then
\begin{align*}
\det W = \det 
\begin{bmatrix}
w_0 & w_1 & \cdots & w_{q-2} & w_{q-1} \\
w_{q-1} & w_{0} & \cdots & w_{q-3} & w_{q-2} \\
\vdots & \vdots & \ddots & \vdots & \vdots \\
w_2 & w_3 & \cdots & w_0 & w_1 \\
w_1 & w_2 & \cdots & w_{q-1} & w_0
\end{bmatrix} = 0
\end{align*}
if and only if either $\sum_{0 \leq i \leq q-1} w_i = 0$ or all the $w_i$ are equal.
\end{theorem}
\begin{proof}
We will make use of the notation introduced in the previous theorem here. 
If $\sum_{i} w_i = 0$ then $\lambda_0 = 0$ and so $\det W =0$. Suppose all the $w_i$
are equal then for $\ell > 0$
\begin{align*}
\lambda_{\ell} &= w_0 (1 + \epsilon^{\ell} +\cdots + \epsilon^{\ell(q-1)}) \\
	&= w_0 \left( \frac{\epsilon^{\ell q}-1}{\epsilon-1}\right)\\
	&= 0, \text{ since $\epsilon^q = 1$.}
\end{align*}
Assume that $\det W = 0$ and that $\lambda_0 \neq 0$. Then by the formula for
the determinant $\lambda_{\ell} = 0$ for some $\ell < q$. By the formula for $\lambda_{\ell}$,
$\epsilon^{\ell}$ is a root of the polynomial
\begin{align*}
	p(x) = \sum_{0 \leq i \leq q-1} v_i x^i.
\end{align*}
However, since $q$ is prime, $\epsilon^{\ell}$ is also a primitive $q$-th root of unity,
and the minimal polynomial for a primitive $q$-th over the rationals is the $q$-th cyclotomic polynomial $f(x) = \sum_{0 \leq i \leq q-1} x^i$. 
This is an irreducible polynomial modulo $p$ (by our assumption) and hence is also the minimal polynomial for $\epsilon^{\ell}$ over $\mathbb{F}_p$. Thus $p(x)$ is a constant
multiple of $f(x)$ and thus all $w_i$ are equal.
\qed \end{proof}

\begin{theorem} \label{thm_critinert:this}
Let $q$ be a prime and $f(x) = \sum_{0 \leq i < q} x^i$ be the $q$-th Cyclotomic
polynomial. Then $g(x)$ is irreducible modulo a prime $p$, iff $p \equiv g \mod q$
where $g$ is a generator of the cyclic group $\mathbb{F}_q^*$.
\end{theorem}
\begin{proof}
Let $K = \mathbb{F}_p$. Every root, $r$, of $f(x)$ over the algebraic
closure $\overline{K}$, satisfies $r^q \equiv 1$. In other words,
they are elements of multiplicative order $q$ in $\overline{K}^*$.
The smallest extension $\mathbb{F}_{p^s}$ that contains
elements of multiplicative order $q$ is the smallest $s$ such that $p^s \equiv 1 \mod q$.
This is the order of $p$ in the multiplicative group $\mathbb{F}_{q}^*$. Suppose
$p \equiv g \mod q$ then the smallest such $s$ is $q-1$.
The field $\mathbb{F}_{p^{q-1}}$ is the splitting field of the polynomial $f(x)$
and consequently, $f(x)$ is irreducible in $\mathbb{F}_p$.
Now if the order of $p$ modulo $q$ is $< q-1$. Then, there is
an extension of $\mathbb{F}_p$, (say) $\mathbb{F}_{p^e}$, that contains
a root $\alpha$ of $f(x)$. Now the polynomial $f'(x) = \prod_{\sigma \in \Gal(\mathbb{F}_{p^e}/\mathbb{F}_p)}(x-\sigma(\alpha))$ is a factor of $f(x)$
which has coefficients in $\mathbb{F}_p$ of degree $e < q-1$. Consequently, $f(x)$
is not irreducible over $\mathbb{F}_p$.
\qed \end{proof}

A bit of algebraic number theory gives some more information: The polynomial $f(x)$
is irreducible modulo $p$ iff $p$ remains inert in the $q$-th cyclotomic field $\mathbb{Q}(\mu_q)$,
where $\mu_q = \exp(2\pi \imath/q)$. This happens iff the Artin symbol at $p$ is
a generator of the Galois group $\Gal(\mathbb{Q}(\mu_q))$. By the Chebotarev
density theorem this happens for a constant density of primes, indeed, the
density is $\frac{\varphi(q-1)}{q-1}$.\\

\end{document}